\documentclass[prl,twocolumn,aps,floatfix,showpacs,tightenlines,superscriptaddress,amsmath,amssymb,amsfonts,10pt]{revtex4-1}
\usepackage{amssymb,amsbsy,epsfig,color,graphicx}
\usepackage{color}
\usepackage{physics}
\usepackage{mathtools}

\usepackage[caption=false]{subfig}
\begin{document}

\title{Non-Markovianity-assisted optimal continuous variable quantum teleportation}

\author{Gianpaolo Torre}

\affiliation{Dipartimento di Ingegneria Industriale, Universit\`a degli Studi di Salerno, Via Giovanni Paolo II 132, I-84084 Fisciano (SA), Italy}

\author{Fabrizio Illuminati}

\email[Corresponding author: ]{filluminati@unisa.it}

\affiliation{Dipartimento di Ingegneria Industriale, Universit\`a degli Studi di Salerno, Via Giovanni Paolo II 132, I-84084 Fisciano (SA), Italy}
\affiliation{INFN, Sezione di Napoli, Gruppo collegato di Salerno, I-84084 Fisciano (SA), Italy}

\date{\today}

\begin{abstract}
We study the continuous-variable (CV) quantum teleportation protocol in the case that one of the two modes of the shared entangled resource is sent to the receiver through a Gaussian Quantum Brownian Motion noisy channel. We show that if the channel is engineered in a non-Markovian regime, the information backflow from the environment induces an extra dependance of the phase of the two-mode squeezing of the shared Gaussian entangled resource on the transit time along the channel of the shared mode sent to the receiver. Optimizing over the non-Markovianity dependent phase of the squeezing yields a significant enhancement of the teleportation fidelity. For short enough channel transit times, essentially unit fidelity is achieved at realistic, finite values of the squeezing amplitude for a sufficiently large degree of the channel non-Markovianity.
\end{abstract}


\maketitle

A realistic and fruitful implementation of quantum information technologies has to come to terms with the unavoidable interaction between the system of interest and the surrounding environment, that typically has a detrimental effect on the quantum properties of the resource, leading to decoherence (see e.g. Refs.~\cite{Zurek,1464-4266-7-4-R01,Schlosshauer2005} for reviews).
In recent years, significant efforts have been devoted to an in-depth understanding of the memory effects that arise in the open evolution of a quantum system. Indeed, quantum non-Markovian dynamics appears to play an important role in the understanding the behaviour of a variety of natural and artificial systems, ranging from biological matter~\cite{Lambert2013, Thorwart2009234, Chin2013, doi:10.1080/00405000.2013.829687} to photonic band-gap materials~\cite{6833787}. Moreover, suitable engineering of the environment and its memory effects, combined with the possibility of properly engineering the quantum resources required to the realization of quantum information tasks~\cite{PhysRevA.76.022301, PhysRevA.81.012333, PhysRevA.82.062329, PhysRevA.93.033807}, opens the way to an improvement of the efficiency of quantum technologies, from quantum cryptography~\cite{PhysRevA.83.042321} and quantum metrology~\cite{PhysRevLett.109.233601} to optimal control~\cite{PhysRevA.85.032321, 6870898, 1367-2630-17-6-063031} and superdense coding~\cite{0295-5075-114-1-10005}.

It is thus important to investigate thoroughly the characterization and quantification of the non-Markovianity of quantum channels~\cite{0034-4885-77-9-094001, RevModPhys.88.021002} and their possible applications in the realization of tailored environments allowing for the enhanced performance of quantum technology tasks in realistic conditions.
In this respect, important progress has been achieved in the characterization of non-Markovian open quantum dynamics of infinite-dimensional quantum optical systems. The recently introduced quantifiers of non-Markovianity for Gaussian channels~\cite{PhysRevA.84.052118,PhysRevLett.115.070401,PhysRevA.92.052122,Groblacher(2015)} allow to study the memory effects of the dynamics in connection with other non-classical properties of the evolving quantum systems and their exploitation for quantum information technologies. These results are of relevance because Gaussian states and channels, due to their easily theoretical and experimental manipulation, represent the most widely used resources in the continuous variable (CV) setting of quantum information: their use may range from universal quantum computation~\cite{PhysRevLett.97.110501} to quantum teleportation protocols~\cite{PhysRevLett.80.869, PhysRevA.76.022301, PhysRevA.81.012333, PhysRevA.82.062329,PhysRevA.84.034305}.
In particular, CV quantum teleportation provides a very clean arena for testing strategies of quantum state and process engineering in noisy environments.
So far, investigations have been mainly dedicated to the improvement of the efficiency of this protocol, both in the ideal and in realistic settings, in terms of modified shared entangled resources~\cite{PhysRevA.76.022301, PhysRevA.81.012333, PhysRevA.82.062329}. In parallel, the fast growing development of integrated photonics, that offers new tools for manipulating light in quantum information technologies~\cite{Politi2009, OBrien2009, Thylen2014}, allows for the possibility to teleport quantum states of CV systems on a compact and hence stable architecture. This feature can provide a crucial step towards scalable realization of linear optical quantum computing~\cite{Metcalf2014} and can yield the realization of fully controlled teleportation protocols in engineered environments over the short channel transit times allowed by integration and miniaturization on chip.

Thus motivated, in the present work we investigate the CV quantum teleportation protocol~\cite{PhysRevA.49.1473, PhysRevLett.80.869, PhysRevA.81.012333} in the realistic condition in which the resource mode sent to Bob evolves in a non-Markovian noisy channel. We will show that, due to the channel memory effects, control of the shared entangled resource in terms of the transit time across the channel allows to obtain a significant improvement of the teleportation fidelity, with unit fidelity achievable at finite, realistic values of the squeezing amplitude for a sufficiently short channel transit time and a sufficiently large degree of the channel non-Markovianity. Further, we will clarify in detail the close relation between the phase-optimization of the teleportation fidelity and the channel non-Markovianity in terms of the time behaviour of non-Markovianity measures based on the violation of the divisibility of the open quantum dynamics~\cite{PhysRevLett.115.070401}.

\begin{figure}
\includegraphics*[width=8.5cm]{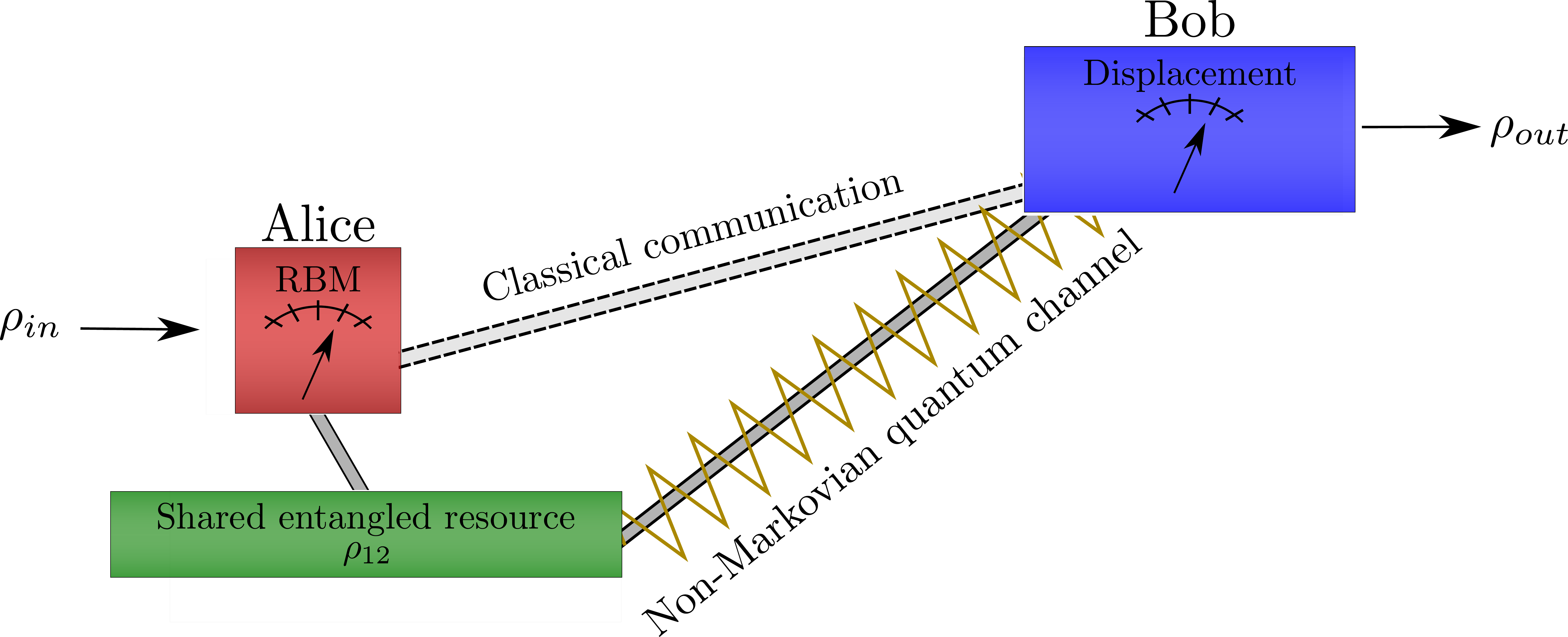}
\caption{(Color online) CV quantum teleportation protocol in the presence of non-Markovian noise. In the first step, Alice mixes the input mode with her mode of the shared entangled resource; the ensuing state is then subject to a realistic Bell measurement. The measurement result is communicated to Bob through a classical communication channel. Depending on the measurement output, Bob applies a unitary transformation to recover the teleported input state. Bob's mode of the shared entangled resource is affected by decoherence during the propagation in the non-Markovian QBM channel. The ensuing output state is the final teleported state. The teleportation fidelity is maximized by fixing the squeezing phase $\phi$ at an optimal value determined by the channel transit time. The non-Markovian regime is engineered experimentally by tuning the ratio between the characteristic frequency $\omega_0$ of the system oscillator and the characteristic cut-off frequency $\omega_c$ of the bath, namely the ratio $x=\omega_c/\omega_0$.}
\label{FigRealNonMarkQuantTel}
\end{figure}

The standard Braunstein-Kimble-Vaidman CV teleportation protocol is schematically illustrated in Fig.~\ref{FigRealNonMarkQuantTel}; a more detailed exposition is provided in the Appendix at the end of the main text~\cite{SuppMat}. The unknown input coherent state is denoted by  $\rho_{in}=\ket{\beta}_{in}\prescript{}{in}{\bra{\beta}}$, with unknown complex amplitude $\beta$. The shared resource mode sent to Bob is affected by a decoherence process during its propagation; it is schematized by a non-Markovian Quantum Brownian Motion (QBM) channel~\cite{PhysRevD.45.2843, PhysRevA.67.042108, PhysRevA.70.032113}, described by the following master equation with time-dependent coefficients:
\begin{align}
\label{QBMEq}
\frac{d\rho}{dt} = & -\frac{i}{\hbar} [H_0,\rho(t)]-\Delta(t)[\hat{x},[\hat{x},\rho(t)]] + \nonumber \\ & +
 \Pi(t)[\hat{x},[\hat{p},\rho(t)]] - i\gamma(t)[\hat{x},\{\hat{p},\rho(t)\}]\;,
\end{align}
where $H_0$ is the fee hamiltonian of the system, the coefficient $\gamma(t)$ represent the damping factor, $\Delta(t)$ and $\Pi(t)$ are the normal diffusion and the anomalous diffusion coefficients respectively, and $\hat{x}$ and $\hat{p}$ are the quadrature operators.
In the following, we will consider the rather typical Ohmic-like spectral density $J(\omega)=\omega e^{-\omega/\omega_c}$, where $\omega_c$ is the cut-off frequency of the bath. In the high-temperature regime, namely the case in which the classical thermal energy is much
larger than the typical one that characterizes the system evolution, one has $K_B T \gg \hbar \omega_c, \hbar \omega_0$, where $\omega_0$ is the characteristic frequency of the system. Focusing on this regime is motivated by the fact that in this case the destructive effects of the system-environment interaction dominate, and any improvement of the efficiency of the protocol in this regime is especially relevant (see for further details the Appendix below~\cite{SuppMat}). At any fixed given temperature and dimensionless time $\tau=\omega_c t$, the {\em non-Markovianity} parameter $x$ that characterizes the scales of the system time evolution identifies with the ratio between the bath and the system frequencies $x=\omega_c/\omega_0$:
for $x\ll 1$, the dynamics is non-Markovian, otherwise it falls in the Markovian regime~\cite{PhysRevA.70.032113, TorreAsym2017}.
The shared Gaussian entangled resource is taken in the class of two-mode squeezed vacuum states $\rho_{12}=S_{12}(\zeta)\ket{00}_{12}\prescript{}{12}{\bra{00}}S_{12}(\zeta)$, where $S_{12}(\zeta)=e^{\zeta a_1^\dag a_2^\dag-\zeta^* a_1 a_2}$ is the two-mode squeezing operator, with $\Re(\zeta)=\cosh(r)$, $\Im(\zeta)=e^{\imath\phi}\cosh(r)$, and $r$ and $\phi$ are, respectively, the amplitude and the phase of the squeezing. As summarized in Eqs.~(18)--(20) of the Appendix below~\cite{SuppMat}, the efficiency of the quantum teleportation protocol depends strongly on the time that the Bob mode spends travelling in the QBM channel (channel transit time). This transit time is defined as the difference $\delta t=t_f-t_0$ between the instant $t_0$  when the resource mode starts to evolve in the channel, and the instant $t_f$ when it reaches Bob. In order to characterize the state evolution in the QBM channel entirely in terms of dimensionless quantities, we introduce the dimensionless channel transit time $\delta\tau=\omega_c \delta t$.

In terms of the parameters that characterize the states and the channel, the teleportation fidelity reads (details are reported in the Appendix below~\cite{SuppMat}):
\begin{align}\label{TelepFid}
&\mathcal{F}(r,\phi,\delta\tau,x) = \frac{1}{\sqrt{ -4\bar{W}_{12}^2(\delta\tau,x)+\dfrac{1}{4} e^{-2\Gamma(\delta\tau,x)} \Lambda_{11} \Lambda_{22}
	}} \, ,
\end{align}
where
\begin{align}
	&\Lambda_{jj}\!=\!\cosh 2r +e^{\Gamma(\delta\tau,x)}\!\!\left(\!\!2\!+\!4\bar{W}_{jj}(\delta\tau,x)\!+2\dfrac{\mathcal{R}^2}{\mathcal{T}^2}\!+\!\cosh 2r\!\!\right)
	\nonumber\\
	& + \qquad\quad 2e^{\Gamma(\delta\tau,x)/2}\cos\left(\phi-\dfrac{\delta\tau}{x}\right)\sinh 2r , \quad j=1,2 \, .
	\label{cofficients}
	\end{align}
In the above,
$\mathcal{T}$ and $\mathcal{R}=\sqrt{1-\mathcal{T}^2}$ are, respectively, the transmissivity and the reflectivity of the beam splitters that model the losses of the realistic Bell measurement performed by Alice and include the effects of the noise affecting Alice's shared resource mode. Finally, $\Gamma(\delta\tau,x)$, $\bar{W}_{11}(\delta\tau,x)$, $\bar{W}_{22}(\delta\tau,x)$, and $\bar{W}_{12}(\delta\tau,x)$ characterize the time evolution~\cite{SuppMat}.
From Eqs.~(\ref{TelepFid}) we note that, at variance with the Markovian case~\cite{PhysRevA.81.012333}, the phase of the squeezing, whose choice contributes
to select the best achievable efficiency of the protocol, acquires an explicit dependance on the channel transit time $\delta\tau$.
Consequently, we introduce the phase-optimized fidelity:
\begin{equation}\label{OptFid}
\mathcal{F}_{\textrm{opt}}(r, \delta\tau, x)=\max_{\phi}\mathcal{F}(r,\phi, \delta\tau, x) \, .
\end{equation}
Due to the positivity of the  master equation coefficients $\bar{W}_{11}(\delta\tau, x)$ and $\bar{W}_{22}(\delta\tau, x)$~\cite{PhysRevA.80.062324},
one finds that at fixed $x$ and $\delta\tau$ the maximum in Eq.~(\ref{OptFid}) is achieved by
\begin{equation}\label{OptPhi}
\phi_{\textrm{opt}}(x,\delta\tau)=\pi+\dfrac{\delta\tau}{x} \, .
\end{equation}
In the Markovian limit $x\gg 1$ we recover the known result~\cite{PhysRevLett.95.150503} that the optimal phase of the squeezing that maximizes the teleportation fidelity is $\phi_{\textrm{opt}} = \pi$, independent of the channel transit time. Therefore a further step has to be included in the preparatory stage of the protocol: the optimal phase $\phi_{\textrm{opt}}$ of the shared squeezed resource has to be selected in the laboratory according to the value of the transit time $\delta \tau$ of Bob's mode in the non-Markovian channel. In turn, the channel transit time is fixed in the experimental setup by the propagation velocity of the shared resource mode along the non-Markovian channel and the length of the latter.

\begin{figure}
	\includegraphics[width=\linewidth]{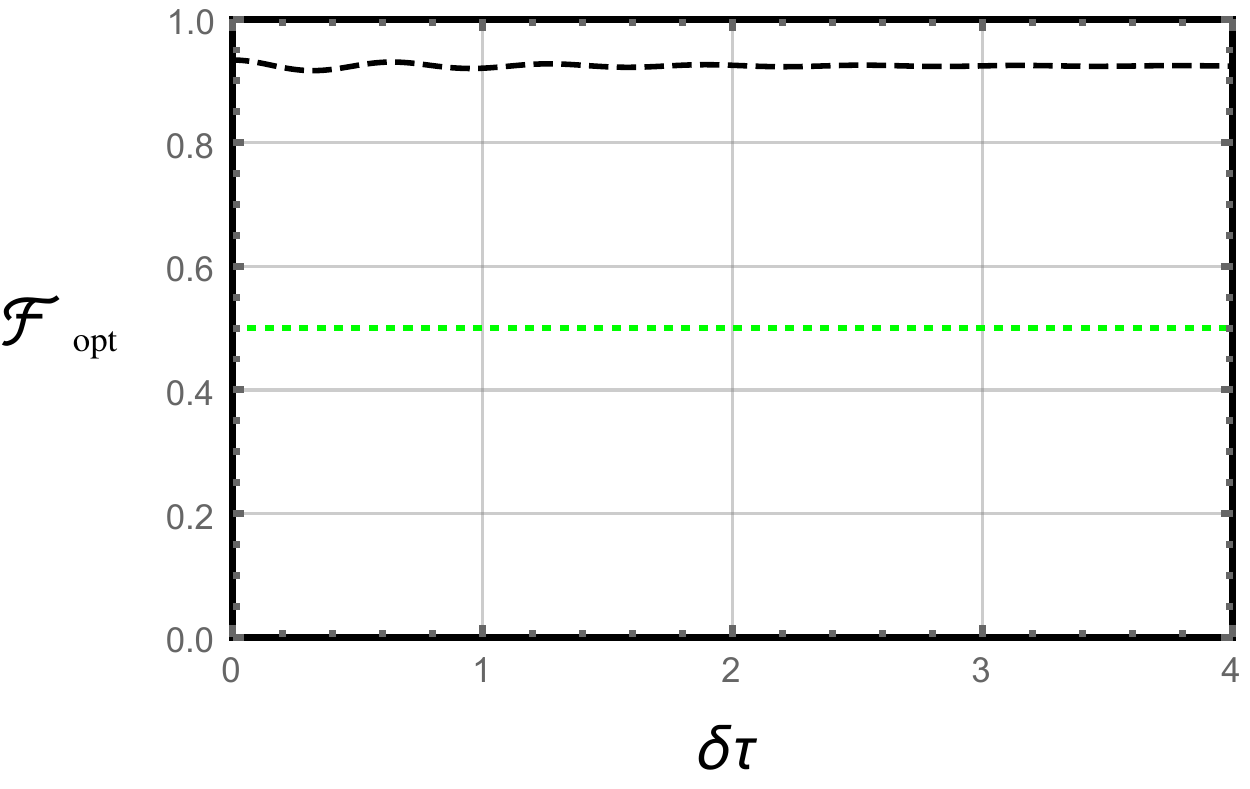}
	\caption{(Color online) Optimal teleportation fidelity Eq.~(\ref{TelepFid}) as a function of the channel transit time in the non-Markovian regime $x=0.1$ (black  dashed curve). The maximum of the fidelity is obtained by choosing the optimal phase Eq.~(\ref{OptPhi}) in Eq.~(\ref{TelepFid}). The classical threshold is also reported for comparison (green dotted horizontal line). The protocol operates in the high-temperature regime $\frac{K_B T}{\hbar \omega_c}=100$. The squeezing amplitude is fixed at $r=2.0$, and the transmissivity at $\mathcal{T}^2=0.90$.}\label{OptFidFid}
\end{figure}

The importance of this optimization strategy is reflected in the behaviour of the optimal fidelity as a function of the channel transit time, as reported in Fig.~\ref{OptFidFid}. Every time interval $\delta \tau$ corresponds to a different experimental setup, namely a different distance between Bob and its resource mode. The squeezing amplitude $r$ of the entangled resource is fixed at $r=2.0$, a value corresponding to the current technological limit~\cite{SchnabelPrivComm}.
In Fig.~\ref{OptFidFid} the black dashed curve represents the optimized fidelity in an intermediate non-Markovian regime $x=0.1$. Further lowering of $x$ and/or increasing of $\mathcal{T}$ at fixed squeezing realizes an essentially unity fidelity for longer channel transit times at lower temperatures.

\begin{figure}
\includegraphics*[width=8.5cm]{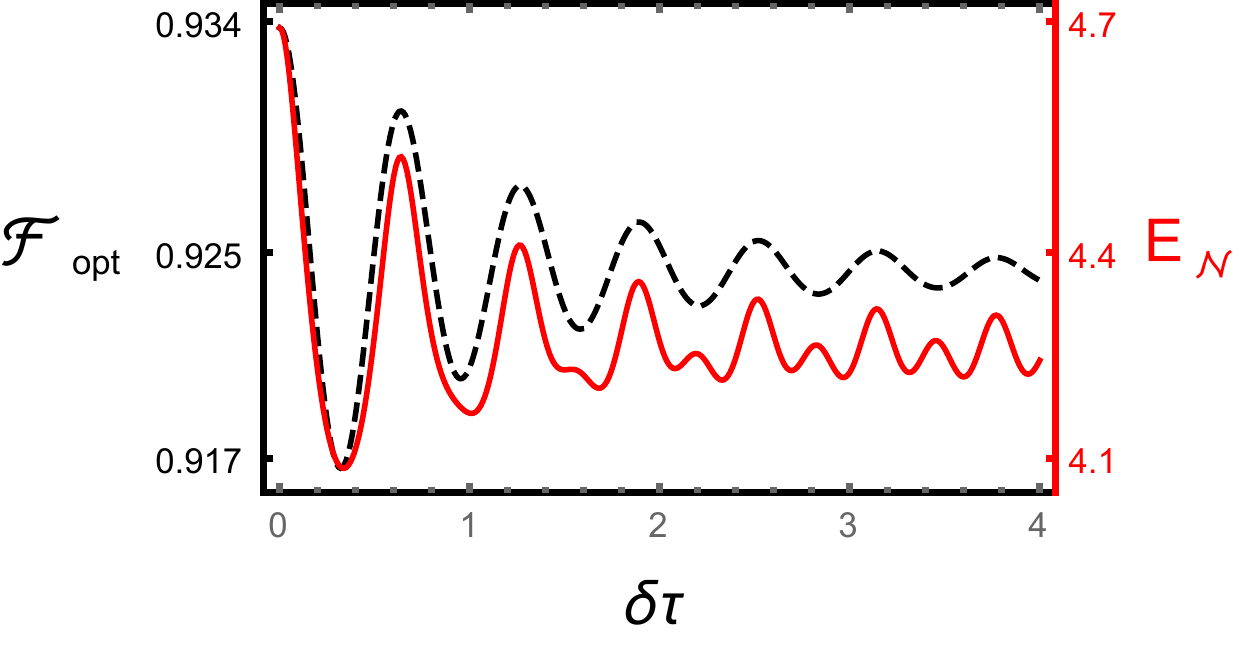}
\caption{(Color online) Optimal teleportation fidelity Eq.~(\ref{OptFid})(black dashed curve) and logarithmic negativity Eq.~(\ref{LogNeg}) (red full curve) as functions of the channel transit time in the non-Markovian regime $x=0.1$. The protocol operates in the high-temperature regime $\frac{K_B T}{\hbar \omega_c}=100$. The squeezing amplitude is fixed at $r=2.0$, and the transmissivity at $\mathcal{T}^2=0.90$.}
\label{FidEnt}
\end{figure}

\begin{figure}
\includegraphics*[width=8.5cm]{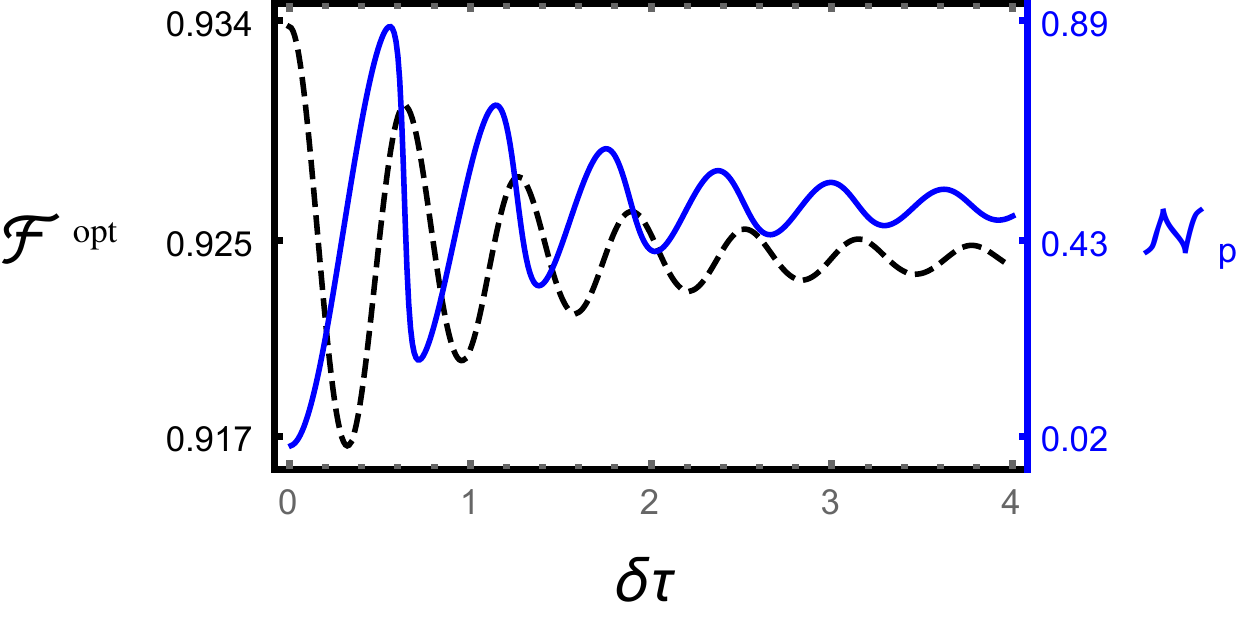}
\caption{(Color online) Optimal teleportation fidelity Eq.~(\ref{OptFid})(black dashed curve) and non-Markovianity measure Eq.~(\ref{FullFormNMMeas}) (blue full line) as functions of the channel transit time in the non-Markovian regime $x=0.1$. The protocol operates in the high-temperature regime $\frac{K_B T}{\hbar \omega_c}=100$. The squeezing amplitude is fixed at $r=2.0$, and the transmissivity at $\mathcal{T}^2=0.90$..}
\label{FidNMark}
\end{figure}

The dependance of the teleportation fidelity on the channel non-Markovianity can be further understood in terms of the entanglement of the shared resource, quantified by the logarithmic negativity $E_{\mathcal{N}}$. Indeed, if one of the shared resource modes is subject to a quantum non-Markovian noise, the entanglement of the shared resource acquires a nontrivial dependance on the channel transit time:
\begin{equation}\label{LogNeg}
E_{\mathcal{N}}(\delta\tau)=\max\{0,-\log\tilde{\nu}_-(\delta\tau)\} \, ,
\end{equation}
where $\tilde{\nu}_-(\delta\tau)$ is the smallest symplectic eigenvalues of the covariance matrix of the state evolved in the channel~\cite{1751-8121-40-28-S01}. The technical details concerning the evaluation of Eq.~(\ref{LogNeg}) are reported in the Appendix below~\cite{SuppMat}. In the ideal, noiseless CV teleportation protocol with shared Gaussian entangled resource the optimal teleportation fidelity, i.e. the one obtained maximizing over all local single-mode operations, is in one-to-one correspondence with the entanglement of the shared resource, as first proved in Ref.~\cite{PhysRevLett.95.150503}. In the presence of Markovian noise the shared entanglement is still the key requirement for a reliable teleportation, but is quickly degraded~\cite{PhysRevA.81.012333}. In the presence of non-Markovian noise, the information backflow from the environment to the system allows for its revival and optimization according to Eq.~(\ref{LogNeg}) above. In Fig.~\ref{FidEnt} we compare the optimized fidelity Eq.~(\ref{OptFid}) and the entanglement Eq.~(\ref{LogNeg}) as functions of the channel transit time $\delta\tau$. We see that indeed, in the presence of a non-Markovian dynamics the one-to-one correspondence between the shared entanglement and the optimal fidelity of the ideal noiseless protocol is essentially recovered.

It is possible to quantify in an even more direct way the crucial role played by a structured bath with suitably engineered non-Markovian noise in the revival of quantum coherence by investigating the non-Markovian character of the QBM channel by appropriate quantifiers. Here we will employ the punctual measure of non-Markovianity $\mathcal{N}_p$ introduced in Ref.~\cite{PhysRevLett.115.070401} and defined in terms of a necessary and sufficient criterion based on the violation of the divisibility of the intermediate quantum dynamical maps~\cite{PhysRevLett.115.070401} (See the Appendix below~\cite{SuppMat} for details on the derivation of $\mathcal{N}_p$ and its properties). For the instance of the QBM channel one has~\cite{TorreAsym2017}:
\begin{equation} \label{FullFormNMMeas}
\mathcal{N}_p \!=\! \dfrac{1}{2}\left[1-\dfrac{\Delta(\delta\tau,x)}{\sqrt{\Delta(\delta\tau,x)^2+\gamma(\delta\tau,x)^2+\Pi(\delta\tau,x)^2}}\right].
\end{equation}
In Fig.~\ref{FidNMark} we report the optimal fidelity Eq.~(\ref{OptFid})and the non-Markovianity measure Eq.~(\ref{FullFormNMMeas}) as functions of the channel transit time $\delta\tau$. At the input time, the fidelity is maximum, corresponding to the ideal decoherence-free case, while the non Markovianity is zero, as the QBM channel coincides with the identity. At later times the fidelity, after a decreasing stage due to the destructive effect of the system-environment interaction, increases, due to the memory effect of the non-Markovian channel, as shown by the behavior of the non-Markovianity measure. An increasing  non-Markovianity of the channel corresponds to an increasing teleportation fidelity, with a time delay. The time shift is due to the fact that the correlation time scale of the environment $\tau_E$ is equal or greater than the relaxation time scale of the system $\tau_S$, corresponding to the rate of change of the state of the system due to the system-environment interaction, quantified by $x=\omega_c / \omega_0 = \tau_S / \tau_E$. The time delay in the transmission to the system of the effect of the interaction with the environment can be well understood in terms of generic collisional models of non-Markovian channels~\cite{PhysRevA.87.040103}.

Some considerations are in order. Non-Markovian effects usually occur when the time evolution satisfies the condition $t \lesssim \tau_E=1/\omega_c$~\cite{PhysRevA.70.032113}. This relation is further constrained in the hight-temperature regime (See the Appendix below~\cite{SuppMat} for details). Indeed, since we have set $k_B T/\hbar\omega_c=10^2$, the characteristic frequency of the bath depends on the temperature through the relation $\omega_c\simeq 10^9 s^{-1} K^{-1} T$. Consequently, for $T\simeq 10^2 K$, the non-Markovian effects are maximally enhanced in the early stage of the evolution. These aspects of the system dynamics are then especially relevant in the case of miniaturized, on-chip quantum information technologies, due to the very small size of the devices and the ensuing very short channel transit times. Important progress in this direction has recently been achieved with the generation of continuous variable Gaussian entangled resource states on a chip~\cite{Masada2015}.
The results of the present investigation encompass a large spectrum of interaction models, since they hold essentially for the entire family of spectral densities $J(\omega)=\omega_c(\frac{\omega}{\omega_c})^se^{-\omega/\omega_c}$, which for $s=1$ include the Ohmic case~\cite{PhysRevA.79.052120}. We have also considered the sub-Ohmic ($s=\frac{1}{2}$) and supra-Ohmic ($s=3$) cases, obtaining qualitatively very similar results.

In summary, we have shown how, by engineering an appropriate time evolution regime, i.e. by fixing the value of the non-Markovian parameter $x$, and adapting accordingly the shared entangled resource, the efficiency of the CV quantum teleportation protocol can be optimized in terms of the phase of the squeezing of the shared resource, yielding a significant improvement of the teleportation fidelity. The present study can be readily generalized to more general cases in several directions, for instance by considering non-Gaussian entangled resources of the squeezed Bell type~\cite{PhysRevA.81.012333}. The present study might be also considered as a step forward towards the construction of a general resource theory of quantum non-Markovianity for quantum information and quantum technology tasks, along the lines outlined in Ref~\cite{BrandaoPRL2015}.

\section{Appendix}

\subsection{Quantum Brownian Motion channel}
In this Section we review the solution of the Quantum Brownian motion (QBM) master equation Eq.~(1) considered in the main text (See also Refs.~\cite{PhysRevD.45.2843, PhysRevA.67.042108, PhysRevA.70.032113} for further details).

We consider the dynamics of a quantum mechanical oscillator with characteristic frequency $\omega_0$ in contact with a bath of harmonic oscillators via a position-position coupling and a factorized initial state.
Under these conditions the system evolution is described by the following master equation:
\begin{align}
\label{HPZEq}
\frac{d\rho}{dt} = & -\frac{i}{\hbar} [H_0,\rho(t)]-\Delta(t)[\hat{x},[\hat{x},\rho(t)]]+
\Pi(t)[\hat{x},[\hat{p},\rho(t)]] \nonumber \\ & -i\gamma(t)[\hat{x},\{\hat{p},\rho(t)\}]\;,
\end{align}
where $\Delta(t)$ and $\Pi(t)$ are, respectively, the normal and anomalous diffusion coefficients, $\gamma(t)$ is the damping coefficient,
and $\hat{x}$ and $\hat{p}$ are the quadrature operators.

In the weak coupling regime, the explicit expressions of the coefficients read:
\begin{align}
\gamma(t)&\!=\!\alpha^2\!\int_0^t  \, ds\int_0^{+\infty} \, d\omega J (\omega)  \sin (\text{$\omega $s}) \sin (\text{$\omega_{0} $s}),\label{gamma} \\
\Delta(t)&\!=\!\alpha^2\!\int_0^t  \,\!\! ds\int_0^{+\infty} \, \!\!\!\!\!\!\!\! d\omega J (\omega) (2N(\omega , T)+1) \cos (\text{$\omega $s}) \cos (\text{$\omega_{0} $s}),\label{delta} \\
\Pi(t)&\!=\!\alpha^2\!\int_0^t  \,\!\! ds\int_0^{+\infty} \, \!\!\!\!\!\!\!\! d\omega J (\omega) (2N(\omega , T)+1) \cos (\text{$\omega $s}) \sin (\text{$\omega_{0} $s}),\label{pi}
\end{align}
where $\alpha \ll 1$ is the coupling constant, $N(\omega , T)=[\exp(\hbar \omega/K_B T)-1]^{-1}$ is the mean photon number, $J(\omega)$ is the spectral density of the bath, that models the system-environment coupling, and $\omega_c$ is the cut-off frequency of the bath.
We work in the hight-temperature regime, namely the case in which the classical thermal energy scale is much larger than the typical ones that characterize the system evolution ($K_B T \gg \hbar \omega_c, \hbar \omega_0$). Indeed, investigating the improvement of the efficiency of the protocol in this regime is particularly interesting, due to the strongly destructive effect of the system-environment interaction at high temperatures. In this regime we can set $2N(\omega, T)+1\approx \frac{2K_B T}{\hbar \omega}$ and determine the explicit expressions of the master equation coefficients for various classes of spectral densities~\cite{PhysRevA.80.062324}.

When considering continuous variable systems it is useful to represent the state $\rho$ in terms of the characteristic function $\chi(\rho)[\Lambda]=\textrm{Tr}[\rho D(\Lambda)]$, where $D(\Lambda)=\exp[i S^\intercal \Omega\Lambda]$ is the displacement operator,
\begin{equation}
\Omega=\left(\begin{array}{cc}\label{SympMat}
0 & 1 \\
-1 & 0
\end{array} \right)
\end{equation}
is the symplectic matrix, $S=(\hat{x},\hat{p})^\intercal$ is the vector of quadrature operators, and $\Lambda=(x,p)^\intercal$ is the vector of coordinates. In the characteristic function description the solution of the master equation Eq.~(\ref{HPZEq}) is~\cite{PhysRevA.67.042108, PhysRevA.70.032113}:
\begin{equation}\label{chiEvol}
\chi(\Lambda, t)=\chi(e^{-\frac{\Gamma(t)}{2}}R^{-1}(t)\Lambda, 0)e^{-\Lambda^\intercal \bar{W}(t)\Lambda},
\end{equation}
where:
\begin{align}
&\bar{W}(t)\!=\![R^{-1}(t)]^\intercal\!\!\left[e^{-\Gamma(t)}\!\!\!\int_0^t \!\!ds e^{\Gamma(s)}R^\intercal(s) M(s) R(s)\right]\!\!R^{-1}(t), \label{WBAR}  \\
& M(s)\!=\!\left(\begin{array}{cc}
\Delta(s)& -\Pi(s)/2 \\
-\Pi(s)/2 & 0
\end{array} \right), \\
& \\ \nonumber
& R(t)=\left(\begin{array}{cc}
\cos(\omega_0 t)& \sin(\omega_0 t) \\
-\sin(\omega_0 t)& \cos(\omega_0 t)
\end{array} \right)\label{WBarMat}
\end{align}
and where we have defined
\begin{equation}
\Gamma(t)=2\int_0^t \gamma(s)ds \, .
\end{equation}

\subsection{Continuous variable quantum teleportation under non-Markovian noise}

In this Section we generalize the continuous variable (CV) quantum teleportation protocol of Ref.~\cite{PhysRevA.81.012333} to the case in which the resource mode shared by Bob evolves in a structured QBM channel with memory effects. Let us denote by $\rho_{\textrm{in}}=|\phi\rangle_{\textrm{in}}\;_{\textrm{in}}{\langle\phi|}$ and $\rho_{\textrm{res}}=|\psi\rangle_{12}\;_{12}{\langle\psi|}$ the projectors corresponding, respectively, to the single-mode input state and with the two-mode entangled resource. The total initial state
$\rho_{0}=\rho_{\textrm{in}} \otimes \rho_{\textrm{res}}$ corresponds to the following characteristic function:
\begin{align}
&\chi_{0}(x_{\textrm{in}},p_{\textrm{in}};x_{1},p_{1};x_{2},p_{2})= \nonumber \\
&\;\; =\textrm{Tr}[\rho_{0} \;
D_{\textrm{in}}(x_{\textrm{in}},p_{\textrm{in}})\, D_{1}(x_{1},p_{1})\, D_{2}(x_{2},p_{2})]
\nonumber \\
&\;\;\;\;= \chi_{\textrm{in}}(x_{\textrm{in}},p_{\textrm{in}})\; \chi_{\textrm{res}}(x_{1},p_{1};x_{2},p_{2})  \,,
\label{globcharfuncinitial}
\end{align}
where $D_{j}(x_{j},p_j)$ denotes the displacement operator for the mode $j$ ($j=\textrm{in},1,2$), $\chi_{\textrm{in}}(x_{\textrm{in}},p_{\textrm{in}})$ is the characteristic function of the input state, and $\chi_{\textrm{res}}(x_{1},p_{1};x_{2},p_{2})$ is the characteristic function of the entangled resource.

Referring to the schematic representation of Fig.(1) in the main text, the first step of the protocol consists in a non-ideal Bell measurement performed by Alice, that is, an homodyne measurements on the first quadrature of the mode $1$ and on the second quadrature of the mode $\textrm{in}$.
After such non-ideal Bell measurement, the remaining mode $2$ is left in a mixed state (see Ref.~\cite{PhysRevA.81.012333} for details). The result is then communicated by Alice to Bob through a classical channel whose gain factor is $g$~\cite{PhysRevA.81.012333}.
The mode $2$ of the resource is sent to Bob through the QBM channel Eq.~(\ref{HPZEq}), the state evolution being described by Eq.~(\ref{chiEvol}). After receiving the mode 2, Bob finally performs on it a displacement operation.

As discussed in the main text, in order to describe the evolution of the Bob mode in the QBM channel it is convenient to introduce the dimensionless time $\tau=\omega_c t$ and the non-Markovianity parameter $x=\omega_c/\omega_0$, where $\omega_0$ and $\omega_c$ are respectively the characteristic frequency of the system evolving in the channel and the characteristic frequency characterizing the time scale of the bath. Finally, let us introduce the channel transit or travel time $\delta t=t_f-t_0$, given by the time interval that the quantum system spends travelling in the QBM channel. The quantities $t_0$ and $t_f$ are respectively the initial and the final time of the evolution. Without loss of generality from now on we set $t_0=0$. The dimensionless channel transit time is then defined as $\delta\tau = \omega_c\delta t$.

Following Ref.~\cite{PhysRevA.81.012333}, it is straightforward to compute the characteristic function associated with the output teleported state:
\begin{widetext}
	\begin{align}
	\label{EQCHIFIN}
	\chi_{\textrm{out}}(x_{2},p_{2},\delta\tau) & =  \chi_{\textrm{in}}(g \mathcal{T} x_{2}, g \mathcal{T} p_{2})\times\nonumber \\
	& \chi_{\textrm{res}}\left( g \mathcal{T} x_{2},-g \mathcal{T} p_{2};e^{-\frac{\Gamma(\delta\tau)}{2}}
	\left[x_{2}\cos(\dfrac{\delta\tau}{x})-p_{2} \sin(\dfrac{\delta\tau}{x})\right],e^{-\frac{\Gamma(\delta\tau)}{2}}
	\left[x_{2}\sin(\dfrac{\delta\tau}{x})+p_{2}\cos(\dfrac{\delta\tau}{x})\right]\right) \times \nonumber \\
	& \exp\left[-\left(\bar{W}_{11}(\delta\tau)+\frac{g^{2} \mathcal{R}^{2}}{2}\right)x_{2}^{2}-
	\left(\bar{W}_{22}(\delta\tau)+\frac{g^{2} \mathcal{R}^{2}}{2}\right)p_{2}^{2}-2 \bar{W}_{12}(\delta\tau)x_{2}p_{2}\right] \, ,
	\end{align}
\end{widetext}
where $\bar{W}_{11}(\delta\tau)$, $\bar{W}_{22}(\delta\tau)$, and $\bar{W}_{21}(\delta\tau)=\bar{W}_{12}(\delta\tau)$ are the components of the $\bar{W}$ matrix Eq.~(\ref{WBAR}), $\mathcal{T}$ and $\mathcal{R}$, with $\mathcal{T}^2 = 1 - \mathcal{R}^2$, are, respectively, the transmissivity  and the reflexivity coefficients, that account for the non-ideal Bell measurement performed by Alice~\cite{PhysRevA.81.012333}. The characteristic function for the ideal protocol is recovered in the limits $\mathcal{R} \rightarrow 0$ $(\mathcal{T} \rightarrow 1)$, $g \rightarrow 1$ and $\delta\tau \rightarrow 0$ $(\Gamma(\delta\tau) \rightarrow 0)$~\cite{PhysRevA.76.022301,PhysRevA.81.012333}. 

From Eq.~(\ref{EQCHIFIN}), the teleportation fidelity in the characteristic function description takes the following expression~\cite{Marian2006}:
\begin{equation}
\mathcal{F}(\delta\tau) =
\frac{1}{2\pi} \int dx_2 dp_2 \; \chi_{\textrm{in}}(x_2, p_2) \chi_{\textrm{out}}(-x_2, -p_2,\delta\tau) \, .
\label{Fidelitychi}
\end{equation}
Due to the Gaussian nature of the unknown input coherent state, of the shared entangled resource, and of the QBM channel that preserves the Gaussian character of the input states, it is useful to provide a description of the protocol, considering both the operations performed by Alice and Bob and the dynamics of the entangled mode $2$ in the channel, in terms of the corresponding transformations on the covariance matrices.

Given the input coherent state $\rho_{\textrm{in}}=\ket{\beta}_{\textrm{in}}\prescript{}{\textrm{in}}{\bra{\beta}}$; the two-mode vacuum state $\rho_{12}=S_{12}(\zeta)\ket{00}_{12}\prescript{}{12}{\bra{00}}S_{12}(\zeta)$ that constitutes the shared entangled resource, where $S_{12}(\zeta)=e^{\zeta a_1^\dag a_2^\dag-\zeta^* a_1 a_2}$ is the two-mode squeezing operator, $\Re(\zeta)=\cosh(r)$, $\Im(\zeta)=e^{\imath\phi}\cosh(r)$, with $r$ and $\phi$ being, respectively, the amplitude and the phase of the squeezing; and the $2 \times 2$ identity matrix $\mathbb{1}_2$, the corresponding covariance matrices read~\cite{ferraroLib}:
	\begin{align}
	& \sigma_{\textrm{in}}=\frac{1}{2}\mathbb{1}_2 \, , \label{covin} \\ & \sigma_{\textrm{res}}(r,\phi)=\dfrac{1}{2}\left(\begin{array}{cc}
	A(r)	& C(r,\phi) \\
	C(r,\phi) & A(r)
	\end{array} \right) \, , \label{TMRES}
	\end{align}
with:
\begin{align}
	 &A(r)=\cosh(2 r)\mathbb{1}_2 \, , \label{A0R} \\
	 &C(r,\phi)=\sinh(2 r)\left(
	 \begin{array}{cc}
	 \cos (\phi ) & \sin (\phi ) \\
	 \sin (\phi ) & -\cos (\phi ) \\
	 \end{array}
	 \right) \, . \label{C0RPhi}
\end{align}
Finally, collecting Eq.~(\ref{chiEvol}) and Eq.~(\ref{EQCHIFIN}), and Eqs.~(\ref{covin})--(\ref{C0RPhi}), the covariance matrix of the output state reads:
\begin{widetext}
	\begin{equation}\label{SingmaOut}
	\sigma_{\textrm{out}}(r,\phi,\delta\tau)\!\!=\!\!\left\lbrace
	\frac{1}{2} e^{-\Gamma(\delta\tau) }\!\! \left[e^{\Gamma(\delta\tau) } g^2 \left(\mathcal{T}^2 \cosh (2 r)\!+\!2 	\mathcal{R}^2\!+\!\mathcal{T}^2\right)\!+\!2 e^{\frac{\Gamma(\delta\tau)}{2}} g \mathcal{T} \sinh (2 r) \cos\left(\phi\!-\!\frac{\tau}{x}\right)\!+\!\cosh (2 r)\right]
	\right\rbrace\mathbb{1}_2+2\bar{W}(\delta\tau) \, .
	\end{equation}
\end{widetext}
It is now straightforward to obtain the teleportation fidelity from Eq.~(\ref{SingmaOut}). Indeed, due to the
Gaussian nature of the input and output states, Eq.~(\ref{Fidelitychi}) can be written as:
\begin{equation}\label{GaussFid}
\mathcal{F}(r,\phi,\delta \tau)=\dfrac{1}{\sqrt{\det[\sigma_{\textrm{in}}+\sigma_{\textrm{out}}(r,\phi,\delta\tau)]}}.
\end{equation}
Replacing Eq.~(\ref{SingmaOut}) in Eq.~(\ref{GaussFid}) we finally obtain Eq.~(2) and Eq.~(3) of the main text.

\subsection{Entanglement of the shared resource}

In this Section we compute the entanglement of the shared entangled two-mode squeezed resource and its dependence on the non-Markovian noise affecting it.
Given the noisy dynamics Eq.~(\ref{chiEvol}), the evolution of the resource state, in the characteristic function description, is given by:
\begin{equation}\label{TMChiEvol}
	\chi(X,t)=\chi\left[(\mathbb{1}_2\oplus R^{-1}(t))X,0\right]e^{-X^\intercal(0_2\oplus\bar{W})X} \, ,
\end{equation}
where $X=(x_1,p_1,x_2,p_2)^\intercal$ is the vector of coordinates and $0_2$ is the $2\times 2$ null matrix. Due to the Gaussian nature both of the state and the evolution, the dynamics Eq.~(\ref{TMChiEvol}) can be expressed as a transformation on the covariance matrix of the input state Eq.~(\ref{TMRES})~\cite{EisWolf(2007)}. One has:
\begin{widetext}
\begin{equation}\label{SigmaResEvol}
	\sigma_{\textrm{res}}(r,\phi,\delta\tau)\!\!=\!\!\left[\mathbb{1}_2\oplus e^{-\frac{\Gamma(\delta\tau)}{2}}R^{-1}(\delta\tau)\right]^\intercal\!\!\sigma_{\textrm{res}}(r,\phi)\left[\mathbb{1}_2\oplus e^{-\frac{\Gamma(\delta\tau)}{2}}R^{-1}(\delta\tau)\right]\!\!+\!\!2\left[0_2\oplus\bar{W}(\delta\tau)\right] \, ,
\end{equation}
\end{widetext}
where we have expressed all quantities in terms of the dimensionless channel transit time $\delta \tau$.
It is useful to rewrite Eq.~(\ref{SigmaResEvol}) as:
\begin{equation}\label{SigEvolt}
\sigma_{\textrm{res}}(r,\phi,\delta\tau)=\left(\begin{array}{cc}
A(r) & C(r,\phi,\delta\tau)  \\
C(r,\phi,\delta\tau) & D(r,\phi,\delta\tau)
\end{array} \right) \, ,
\end{equation}
where $A(r)$ is the matrix defined in Eq.~(\ref{A0R}), and:
\begin{align}
	& C(r,\delta\tau)=e^{-\frac{\Gamma(\delta\tau)}{2}}C(r,\phi)R^{-1}(\delta\tau) \, ,\\
& \nonumber \\
	& D(r,\delta\tau)=e^{-\Gamma(\delta\tau)}A(r)-2\bar{W}(\delta\tau) \, .
\end{align}
In the above, $\bar{W}(t)$ and $C(r,\phi)$ are, respectively, the matrices defined in Eq.~(\ref{A0R}) and Eq.~(\ref{C0RPhi}).
In order to quantify the entanglement of the shared resource state we resort to the logarithmic negativity, Eq.~(5) in the main text~\cite{1751-8121-40-28-S01}:
\begin{displaymath}
E_{\mathcal{N}}(\delta\tau)=\max\{0,-\log\tilde{\nu}_-(\delta\tau)\} \, ,
\end{displaymath}
where $\tilde{\nu}_-(\delta\tau)$ is the smallest symplectic eigenvalue of the
covariance matrix. Its expression in terms of the submatrices composing the covariance matrix Eq.~(\ref{SigEvolt}) is given by~\cite{1751-8121-40-28-S01}:
\begin{widetext}
	\begin{equation}
	\tilde{\nu}_-(r,\phi,\delta\tau)=\sqrt{\!{\dfrac{1}{2}}\tilde{\Delta}(r,\phi,\delta\tau)-\dfrac{1}{2}\!\sqrt{\tilde{\Delta}(r,\phi,\delta\tau)^2
-4\det[\sigma(r,\phi,\delta\tau)^2]}} \, ,
	\end{equation}
\end{widetext}
with:
\begin{equation}
\tilde{\Delta}(r,\phi,\delta\tau)\!=\!\det[A(r)]+\det[D(r,\phi,\delta\tau)]\!-\!2\det[C(r,\phi,\delta\tau)] \, .
\end{equation}

\subsection{Non-Markovianity measure of the QBM channel}

In this Section we briefly review the non-Markovianity measure used to quantify the non-Markovianity content of the QBM channel. Further details can be found in Refs.~\cite{PhysRevLett.115.070401, TorreAsym2017}. The Gaussian nature of the QBM channel allows to characterize it in terms of two $2 \times 2 $ matrices $(X, Y)$, that act on the covariance matrix $\sigma$ of the input state as follows:
\begin{displaymath}
\sigma(t) = X(t)\sigma(0)X(t)^\intercal+Y(t) \, .
\end{displaymath}
For a Gaussian channel, the condition of Complete positivity (CP) is then expressed as:
\begin{equation}
Y(t)-\dfrac{\imath}{2}\Omega+\frac{\imath}{2}X(t)\Omega X(t)^\intercal \geq 0 \, ,
\end{equation}
where $\Omega$ is the symplectic matrix defined in Eq.~(\ref{SympMat}).
The non-Markovianity of the dynamics can be measured by the violation of the divisibility condition. In turn, the violation of the latter is measured by the amount by the QBM channel violates the CP condition for the intermediate map, namely the map that describes the system evolution between two generic time instants. In Ref.~\cite{PhysRevLett.115.070401} we have shown that in any interval $[t, t+\epsilon]$, violation of the CP condition occurs if and only if
\begin{equation}\label{CPINtMapViol}
Y(t+\epsilon, t)-\dfrac{\imath}{2}\Omega+\frac{\imath}{2}X(t+\epsilon, t)\Omega X(t+\epsilon, t)^\intercal < 0 \, ,
\end{equation}
where the matrices $(X(t+\epsilon, t), Y(t+\epsilon, t))$ are given by the following relations:
\begin{align}
 & X(t+\epsilon, t)=X^{-1}(t+\epsilon, 0)X(t,0) \, , \nonumber \\
 & Y(t+\epsilon, t)=Y(t+\epsilon, 0)-X(t+\epsilon, t)Y(t,0)X^\intercal(t+\epsilon, t) \, .
\end{align}
A natural measure of punctual non-Markovianity, namely the non-Markovianity at any given time $t$, can then be defined in terms of the negative part of the spectrum of the matrix appearing in the l.h.s. of Ineq.~(\ref{CPINtMapViol}) as follows~\cite{PhysRevLett.115.070401,TorreAsym2017}:
\begin{equation}\label{NMarkMeas}
\mathcal{N}_p(t)=\lim_{\epsilon\rightarrow 0^+}\dfrac{\sum_{i=1}^2\left(\vert\lambda_i(t+\epsilon, t)\vert-\lambda_i(t+\epsilon, t)\right)}{\sum_{i=1}^2\vert\lambda_i(t+\epsilon, t)\vert} \, ,
\end{equation}
where $\lambda_i(t+\epsilon, t)$ are the eigenvalues of the matrix on the r.h.s. of Ineq.~(\ref{CPINtMapViol}). The expression of the measure of non-Markovianity, Eq.~(\ref{NMarkMeas}) above, has been computed explicitly in Ref.~\cite{TorreAsym2017} for the QBM channel in terms of the channel coefficients Eqs.~(\ref{gamma})--(\ref{pi}) as functions of the dimensionless channel transit time $\delta \tau$ and is reported in Eq.~(6) of the main text.


\begin{thebibliography}{99}

\bibitem{Zurek}
W. H. Zurek, Decoherence, einselection, and the quantum origins of the classical, Rev. Mod. Phys. {\bf 75}, 715 (2003).

\bibitem{1464-4266-7-4-R01}
A. Serafini, M. G. A. Paris, F. Illuminati, and S. De Siena, Quantifying decoherence in continuous variable systems, J. Opt. B. Quantum Semiclassical Opt. {\bf 7}, R19 (2005).

\bibitem{Schlosshauer2005}
M. Schlosshauer, Rev. Mod. Phys., Decoherence, the measurement problem, and interpretations of quantum mechanics, {\bf 76}, 1267 (2005).

\bibitem{Lambert2013}
N. Lambert, Y.-N. Chen, Y.-C. Cheng, C.-M. Li, G.-Y. Chen, and F. Nori, Quantum biology, Nat. Phys. {\bf 9}, 10 (2013).

\bibitem{Thorwart2009234}
M. Thorwart, J. Eckel, J.H. Reina, P. Nalbach, and S. Weiss, Enhanced quantum entanglement in the non-Markovian dynamics of biomolecular excitons, Chem. Phys. Lett. {\bf 478}, 234 (2009).

\bibitem{Chin2013}
A. W. Chin, J. Prior, R. Rosenbach, F. Caycedo-Soler, S. F. Huelga, and M. B. Plenio, The role of non-equilibrium vibrational structures in electronic coherence and recoherence in pigment–protein complexes, Nat. Phys. {\bf 9}, 113 (2013).

\bibitem{doi:10.1080/00405000.2013.829687}
S. F. Huelga and M. B. Plenio, Vibrations, quanta and biology, Contemp. Phys. {\bf 54}, 181 (2013).

\bibitem{6833787}
U. Hoeppe, C. Wolff, J. K\"{u}chenmeister, J. Niegemann, M. Drescher, H. Benner, and K. Busch, Direct Observation of Non-Markovian Radiation Dynamics in 3D Bulk Photonic Crystals, Phys. Rev. Lett. {\bf 108}, 043603 (2012).

\bibitem{PhysRevA.76.022301}
F. Dell'Anno, S. De Siena, L. Albano, and F. Illuminati, Continuous-variable quantum teleportation with non-Gaussian resources, Phys. Rev. A {\bf 76}, 022301 (2007).

\bibitem{PhysRevA.81.012333}
F. Dell'Anno, S. De Siena, and F. Illuminati, Realistic continuous-variable quantum teleportation with non-Gaussian resources, Phys. Rev. A {\bf 81}, 012333 (2010).

\bibitem{PhysRevA.82.062329}
F. Dell'Anno, S. De Siena, G. Adesso, and F. Illuminati, Teleportation of squeezing: Optimization using non-Gaussian resources, Phys. Rev. A {\bf 82}, 062329 (2010).

\bibitem{Politi2009}
A. Politi, J. C. F. Matthews, M. G. Thompson, and J. L. O’Brien,  Integrated Quantum Photonics, IEEE J. Sel. Topics Quantum Electron. {\bf 15}, 1673 (2009).

\bibitem{OBrien2009}
J. L. O'Brien, A. Furusawa, and J. Vuckovic, Photonic quantum technologies, Nat. Photon. {\bf 3}, 687 (2009).

\bibitem{Thylen2014}
L. Thyl\'{e}n and L. Wosinski, Integrated photonics in the 21st century, Photonics Research {\bf 2}, 75 (2014).

\bibitem{Metcalf2014}
B. J. Metcalf et al., Quantum teleportation on a photonic chip, Nat. Photon. {\bf 8}, 770 (2014).

\bibitem{PhysRevA.93.033807}
L.-Y. Hu, Z. Liao, S. Ma, and M. S. Zubairy, Optimal fidelity of teleportation with continuous variables using three tunable parameters in a realistic environment, Phys. Rev. A {\bf 93}, 033807 (2016).

\bibitem{PhysRevA.83.042321}
R. Vasile, S. Olivares, M. G. Paris, and S. Maniscalco, Continuous-variable quantum key distribution in non-Markovian channels, Phys. Rev. A {\bf 83}, 042321 (2011).

\bibitem{PhysRevLett.109.233601}
A. W. Chin, S. F. Huelga, and M. B. Plenio, Quantum Metrology in Non-Markovian Environments, Phys. Rev. Lett. {\bf 109}, 233601 (2012).

\bibitem{PhysRevA.85.032321}
B. Hwang and H.-S. Goan, Optimal control for non-Markovian open quantum systems, Phys. Rev. A {\bf 85}, 032321 (2012).

\bibitem{6870898}
W. Cui, \textit{11th IEEE International Conference on Control Automation (ICCA)}, 72 (2014).

\bibitem{1367-2630-17-6-063031}
V. Mukherjee, V. Giovannetti, R. Fazio, S. F. Huelga, T. Calarco, and S. Montangero, Efficiency of quantum controlled non-Markovian thermalization, New J. Phys. {\bf 17}, 063031 (2015).

\bibitem{0295-5075-114-1-10005}
B.-H. Liu, X.-M. Hu, Y.-F. Huang, C.-F. Li, G.-C. Guo, A. Karlsson, E.-M. Laine, S. Maniscalco, C. Macchiavello, and J. Piilo, Efficient superdense coding in the presence of non-Markovian noise, Europhys. Lett. {\bf 114}, 10005 (2016).

\bibitem{0034-4885-77-9-094001}
\'{A}. Rivas, S. F. Huelga, and M. B. Plenio, Quantum non-Markovianity: characterization, quantification and detection, Rep. Prog. Phys. {\bf 77}, 094001 (2014).

\bibitem{RevModPhys.88.021002}
H.-P. Breuer, E.-M. Laine, J. Piilo, and B. Vacchini, Colloquium: Non-Markovian dynamics in open quantum systems, Rev. Mod. Phys. {\bf 88}, 021002 (2016).

\bibitem{PhysRevA.84.052118}
R. Vasile, S. Maniscalco, M. G. Paris, H.-P. Breuer, and J. Piilo, Quantifying non-Markovianity of continuous-variable Gaussian dynamical maps, Phys. Rev. A {\bf 84}, 052118 (2011).

\bibitem{PhysRevLett.115.070401}
G. Torre, W. Roga, and F. Illuminati, Non-Markovianity of Gaussian Channels, Phys. Rev. Lett. {\bf 115}, 070401 (2015).

\bibitem{PhysRevA.92.052122}
L. A. M. Souza, H. S. Dhar, M. N. Bera, P. Liuzzo-Scorpo, and G. Adesso, Gaussian interferometric power as a measure of continuous-variable non-Markovianity, Phys. Rev. A {\bf 92}, 052122 (2015).

\bibitem{Groblacher(2015)}
S. Groblacher, A. Trubarov, N. Prigge, G. D. Cole, M. Aspelmeyer, and J. Eisert, Observation of non-Markovian micromechanical Brownian motion, Nat. Commun., {\bf 6}, 7606 (2015).

\bibitem{PhysRevLett.97.110501}
N. C. Menicucci, P. van Loock, M. Gu, C. Weedbrook, T. C. Ralph, and M. A. Nielsen, Universal quantum computation with continuous-variable cluster states, Phys. Rev. Lett. {\bf 97}, 110501 (2006).

\bibitem{PhysRevLett.80.869}
S. L. Braunstein and H. J. Kimble, Teleportation of Continuous Quantum Variables, Phys. Rev. Lett. {\bf 80}, 869 (1998).

\bibitem{PhysRevA.84.034305}
G. He, J. Zhang, J. Zhu, and G. Zeng, Continuous-variable quantum teleportation in bosonic structured environments, Phys. Rev. A {\bf 84}, 034305 (2011).

\bibitem{PhysRevA.49.1473}
L. Vaidman, Teleportation of quantum states, Phys. Rev. A {\bf 49}, 1473 (1994).

\bibitem{SuppMat}
See the relevant subsection in the Appendix.

\bibitem{PhysRevD.45.2843}
B. L. Hu, J. P. Paz, Y. and Zhang, Quantum Brownian motion in a general environment: Exact master equation with nonlocal dissipation and colored noise, Phys. Rev. D {\bf 45}, 2843 (1992).

\bibitem{PhysRevA.67.042108}
F. Intravaia, S. Maniscalco, and A. Messina, Density-matrix operatorial solution of the non-Markovian master equation for quantum Brownian motion, Phys. Rev. A {\bf 67}, 042108 (2003).

\bibitem{PhysRevA.70.032113}
S. Maniscalco, J. Piilo, F. Intravaia, F. Petruccione, and A. Messina, Lindblad- and non-Lindblad-type dynamics of a quantum Brownian particle, Phys. Rev. A, {\bf 70}, 032113 (2004).

\bibitem{TorreAsym2017}
G. Torre and F. Illuminati, Exact non-Markovian dynamics of Gaussian quantum channels: Finite-time and asymptotic regimes, arXiv:1804.03095 (2018).

\bibitem{PhysRevA.80.062324}
R. Vasile, S. Olivares, M. G. Paris, and S. Maniscalco, Continuous-variable-entanglement dynamics in structured reservoirs, Phys. Rev. A {\bf 80}, 062324 (2009).

\bibitem{PhysRevLett.95.150503}
G. Adesso and F. Illuminati, Equivalence between entanglement and the optimal fidelity of continuous variable teleportation, Phys. Rev. Lett. {\bf 95}, 150503 (2005).

\bibitem{SchnabelPrivComm}
U. L. Andersen, T. Gehring, C. Marquardt, and G. Leuchs, 30 years of squeezed light generation, Phys. Scr. {\bf 91}, 053001 (2016), and references therein.

\bibitem{1751-8121-40-28-S01}
G. Adesso and F. Illuminati, Entanglement in continuous-variable systems: recent advances and current perspectives, J. Phys. A {\bf 40}, 7821 (2007).

\bibitem{PhysRevA.87.040103}
F. Ciccarello, G. M. Palma, and V. Giovannetti, Collision-model-based approach to non-Markovian quantum dynamics, Phys. Rev. A {\bf 87}, 040103(R) (2013).

\bibitem{PhysRevA.79.052120}
J. Paavola, J. Piilo, K.-A. Suominen, and S. Maniscalco, Environment-dependent dissipation in quantum Brownian motion, Phys. Rev. A {\bf 79}, 052120 (2009).

\bibitem{Masada2015}
G. Masada, K. Miyata, A. Politi, T. Hashimoto, J. L. O'Brien, and A. Furusawa, Continuous-variable entanglement on a chip, Nat. Photon. {\bf 9}, 316 (2015).

\bibitem{BrandaoPRL2015}
F. G. S. L. Brandao and G. Gour, The general structure of quantum resource theoris, Phys. Rev. Lett. {\bf 115}, 070503 (2015).

\bibitem{Marian2006}
P. Marian and T. A. Marian, Continuous-variable teleportation in the characteristic-function description, Phys. Rev. A {\bf 74}, 042306 (2006).

\bibitem{ferraroLib}
A. Ferraro, S. Olivares, and M. G. A. Paris, \textit{Gaussian states in quantum information} (BIBLIOPOLIS, Naples, 2005).

\bibitem{EisWolf(2007)}
J. Eisert and M. M. Wolf, \textit{Quantum Information with Continuous Variables of Atoms and Light}, edited by N. J. Cerf, G. Leuchs, and E. S. Polzik, pages 23-42 (World Scientific, 2007).

\end{thebibliography}
\end{document}